# Ultra-compact nonvolatile phase shifter based on electrically reprogrammable transparent phase change materials

Carlos Ríos[1,2,3,†,*], Qingyang Du[3,†,*], Yifei Zhang[3], Cosmin-Constantin Popescu[3], Mikhail Y. Shalaginov[3], Paul Miller[4], Christopher Roberts[4], Myungkoo Kang[5], Kathleen A. Richardson[5,6], Tian Gu[3,7], Steven A. Vitale[4,*], Juejun Hu[3,7,*]

[1]Department of Materials Science & Engineering, University of Maryland, College Park, MD, USA
[2]Institute for Research in Electronics and Applied Physics, University of Maryland, College Park, MD, USA
[3]Department of Materials Science & Engineering, Massachusetts Institute of Technology, Cambridge, MA, USA
[4]Lincoln Laboratory, Massachusetts Institute of Technology. Lexington, MA, USA
[5]CREOL, The College of Optics & Photonics, University of Central Florida, Orlando, FL USA
[6]Department of Materials Science and Engineering, University of Central Florida, Orlando, FL USA
[7]Materials Research Laboratory, Massachusetts Institute of Technology, Cambridge, MA, USA

[†]These authors contributed equally
*riosc@umd.edu, qydu@xmu.edu.cn, steven.vitale@ll.mit.edu

## Abstract

**Energy-efficient programmable photonic integrated circuits (PICs) are the cornerstone of on-chip classical and quantum optical technologies.[1,2] Optical phase shifters constitute the fundamental building blocks that enable these programmable PICs. Thus far, carrier modulation and thermo-optical effect are the chosen phenomena for ultrafast and low-loss phase shifters, respectively; however, the state and information they carry are lost once the power is turned off—they are volatile. The volatility not only compromises energy efficiency due to their demand for constant power supply, but also precludes them from emerging applications such as in-memory computing. To circumvent this limitation, we introduce a phase shifting mechanism that exploits the nonvolatile refractive index modulation upon structural phase transition of $Sb_2Se_3$, a bi-state transparent phase change material (PCM). A zero-static power and electrically-driven phase shifter was realized on a foundry-processed silicon-on-insulator platform, featuring record phase modulation up to 0.09 $\pi$/µm and a low insertion loss of 0.3 dB/$\pi$, which can be further improved upon streamlined design. We also pioneered a one-step partial amorphization scheme to enhance speed and energy efficiency of PCM devices. A diverse cohort of programmable photonic devices was demonstrated based on the ultra-compact PCM phase shifter.**

The ability to actively control the phase of light is the common denominator in applications ranging from optical switching in data communications[3], beam steering,[4] spectroscopy,[5] optical neural networks,[6] as well as quantum processing and computing with photons.[7] Phase modulation has been achieved with carrier injection or depletion in semiconductors,[8] via electro-optic (EO) effects in liquid crystals,[9] EO polymers,[10] and Pockels crystals,[11] and through the thermo-optic (TO) effect,[12] and mechanical displacement.[13] The small refractive index perturbation from these mechanisms typically results in a large device footprint, often hundreds of microns or larger (Table 1). Other limitations include large optical loss penalty (carrier injection/depletion) and long switching time (liquid crystals), as well as challenges associated with standard foundry process integration (EO polymers, Pockels crystals, and mechanical displacement). Moreover, all the mechanisms above are volatile, mandating a constant power supply to operate. Such volatility significantly increases energy consumption, especially for applications where reconfiguration operations occur at millisecond intervals or even longer. Nonvolatile phase shifters for PICs have been achieved through charge-trapping effects,[14] mechanically latched micro-electromechanical systems,[15] and ferroelectric domain switching,[16] which thus far have suffered from high losses, severe long-term drift, large form factors, or incompatibility with standard Si processes. A compact, low-loss and nonvolatile optical phase shifter (OPS) compatible with Si photonic integration is therefore much coveted.

Chalcogenide phase change materials (PCMs) offer a promising solution to nonvolatile programmable PICs.[17–19] Traditional PCM devices, however, cannot provide phase-only modulation due to high optical losses associated with classical PCMs exemplified by $Ge_2Sb_2Te_5$.[20,21] Moreover, most PCM device prototypes to date still rely upon furnace annealing or an external laser stimulus to trigger the structural transition. This limitation has motivated the development of low-loss PCMs in recent years such as $Ge_2Sb_2Se_4Te$ (GSST),[22] $Ge_3Sb_2Te_6$,[23] $Sb_2S_3$,[24,25] and $Sb_2Se_3$.[26] In particular, $Sb_2Se_3$ offers vanishingly small losses at 1550 nm and an index contrast of $\Delta n \approx 0.77$,[26] making it an ideal PCM for programmable photonics in the telecommunication bands.[27,28] Capitalizing on the unique attributes of $Sb_2Se_3$, this work demonstrates an OPS that simultaneously confers low insertion loss, small form factor, and zero-static power consumption. The OPS devices were fabricated in silicon-on-insulator (SOI) where silicon acts both as the light guide and resistive heater to actuate structural phase transition via single electrical pulses. Notably, the devices were processed on 200-mm wafers leveraging a 90-nm CMOS foundry, enabling them to be seamlessly integrated with industry-standard PIC platforms.

**Results**

The PCM-based phase shifter comprises a 30 nm thick $Sb_2Se_3$ cell, initially prepared in the crystalline state, resting on top of a 220 nm SOI waveguide with a 110 nm rib. This waveguide geometry is designed for transverse electric mode propagation and has been used in all the programmable photonic devices in this work. A section of the waveguide was phosphorus doped to act as a microheater (Fig. 1a). More details of the foundry fabrication process are described in the Methods and the Supplementary Section S1. An essential aspect of the microheater design is the choice of doping concentration given the trade-off between resistivity and optical losses (Fig. 1b). Lower resistivity reduces the pulse voltages required to reach the transition temperatures of $Sb_2Se_3$ at the expense of exacerbated free carrier absorption. After testing four different doping combinations, we found a good balance with a concentration of $n \sim 4 \times 10^{18}$ cm$^{-3}$, which corresponds to an insertion loss of ~ 0.01 dB/μm. Details on the design and optimization of the microheater are described in Supplementary Section S2. Figure 1c shows a scanning electron microscope image of a phase shifter with a 10 μm-long bowtie-shaped doping region. With a refractive index contrast of $\Delta n \approx 0.77$ upon switching (Fig. 1d), mode simulation indicates a change in effective refractive index of $\Delta n_{eff} \approx 0.071$ and a phase shift of 0.09 π/μm at 1565 nm. Given the flat refractive index of Sb$_2$Se$_3$ between 1500-1600 nm range, the phase modulation in this range would only depend on the wavelength and lie within 0.089-0.094 π/μm. This results in an ultra-compact OPS of only 11 μm in length needed to attain full π phase delay, representing significant improvement over the state-of-the-art. The small device footprint also yields a low theoretical insertion loss of 0.01 dB/μm resulting from free carrier absorption. Experimentally, we measured a total insertion loss of 0.03 dB/μm. The loss is larger than the expected value due to other effects such as scattering. This loss change associated with Sb$_2$Se$_3$ crystallization is experimentally assessed to be 0.018 dB/μm, corresponding to a low 0.2 dB/π excess loss, including 0.1 dB/π scattering loss which can be circumvented with improved designs (Supplementary Section S5 and S6).

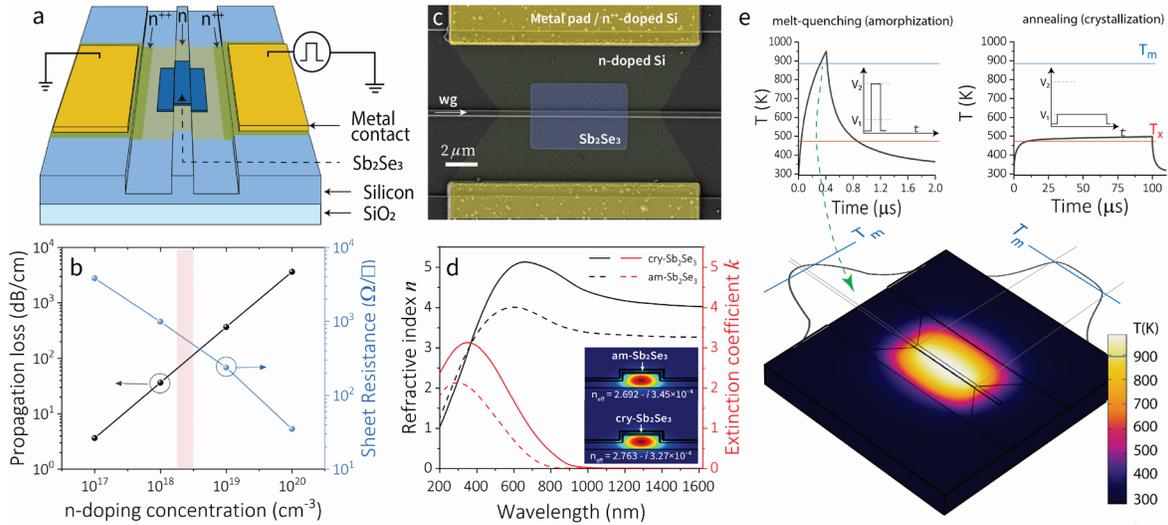

**Fig. 1.** Sb$_2$Se$_3$-based phase shifter using a doped-silicon microheater. **a** Microheater design with doped-silicon profiles to create an electrical resistor in a half-etched 220 nm SOI waveguide. **b** FEM simulated propagation loss and sheet resistivity as a function of phosphorous doping concentration. The highlighted region corresponds to the range where our proposed optimum doping concentration lies. **c** SEM image of a fabricated 6 μm-long bowtie microheater with a 6 μm-long Sb$_2$Se$_3$. **d** Sb$_2$Se$_3$ refractive index in both amorphous and crystalline states with simulated optical modes for each state at 1565 nm wavelength. **e** FEM simulated transient temperature during and after the pulse excitation for both amorphization and crystallization. The three-dimensional (3D) plot shows the temperature profile at the end of the amorphization pulse, reaching the melting temperature of Sb$_2$Se$_3$.

To reversibly switch the device in Fig. 1b, we use Joule heating induced by a voltage pulse that elevates the temperature of the central area of the microheater, which coincides with the waveguide, following similar phase transition dynamics demonstrated in Ref. [29]. The pulse width and voltage depend on the microheater geometry. The high impedance of the microheaters (~ 1 kΩ) led to an approximated 1.94-fold higher voltage dropped on the devices than the voltage set on the pulse generator, which assumes a 50 Ω load. The latter value will be quoted in brackets for reference. For the bowtie geometry in Fig. 1c, which we will extensively use throughout this work for $\pi/2$ phase shift, we used 6.2 V [3.2 V] (3 mA) pulses of varying duration (up to 1 ms) to crystallize the device by heating to above the $Sb_2Se_3$ crystallization temperature, $T_x$ = 200 °C. To amorphize, a single 21 V [10.8 V] (7.8 mA) × 400 ns pulse was employed to raise the temperature over the melting point, $T_m$ = 620 °C (893 K). The crystallization pulse length correlates with the total phase shift; while we observed crystallization with 5 µs pulses, short pulses would induce smaller hotspots with $T > T_x$, crystallizing only small areas of $Sb_2Se_3$. To compensate this effect, we used longer pulses in the 0.1-1 ms range to heat up the entire cell. The large voltage required to amorphize results from the nonlinear I-V response of microheaters, given the strong dependence of the carrier mobility on temperature (see Supplementary Section S2). Figure 1e shows a finite-element method (FEM) simulation of the temperature evolution for both the amorphization and crystallization pulses as well as the 3D temperature profile at the end of the amorphization pulse. Given the large thermal conductivity of silicon, heat rapidly dissipates after the pulse excitation with a decay time of 400 ns, allowing the fast quench process required to amorphize $Sb_2Se_3$. We found that the most prolonged pulse duration able to amorphize was around 1 µs, beyond which crystallization cannot be completely suppressed during cooling (Supplementary Section 2).

We initially demonstrate phase modulation using an unbalanced Mach-Zehnder interferometer (MZI) with a 6 µm-long $Sb_2Se_3$ OPS in each arm. Figure 2b shows the four permutations for the extreme phase states corresponding to a total phase shift of nearly $\pi$. We attribute the difference in extinction ration to the small excess loss upon crystallization, as we discuss below, and the difference between am/am and cry/cry states to small variations in the area that is switched. Continuous (multi-state) switching by partially crystallizing amorphous $Sb_2Se_3$ and a MZI with four permutations and phase modulation exceeding $1.6\pi$ with a longer OPS are also realized and discussed in the Supplementary Section S8. We further demonstrated a 2 × 2 switch – a fundamental building block for constructing programmable photonic architectures[1] – in which light can be switched between two outputs. Since using two 50/50 directional couplers (DC) means that switching between output ports is achieved with either both OPS in phase or with a $\pi$ phase difference, we opted for "unbalanced" directional couplers that display a total equal splitting at the switch output with both OPS in phase. This is achieved by using ~80/20 splitting in each DC, and as result, we could switch between output ports by applying ± $\pi/2$ phase shift in both arms. This is clearly not an optimized MZI switch (design that is beyond the scope of this paper) but it allows us to visualize the resulting amplitude modulation by actuating the $Sb_2Se_3$ phase shifter. Using a 10 µm-long heater with 6 µm-long $Sb_2Se_3$, we show 250 switching events (125 complete cycles) of the MZI (Fig. 2c). Reversible and reliable switching of each phase shifter allows the four different 'full' phase permutations, including splitting equally or directing the light to each of the two output channels with up to 15 dB extinction ratio

(Fig. 2d). An unbalanced directional coupler causes the difference in extinction ratio (ER) between the two opposite phase states at the output of the MZI, but the π/2 phase shift is verified in each arm (Supplementary Section S3). In Fig. S4, we further demonstrate 22 dB extinction ratio for a similar 2 × 2 MZI switch in time domain measurements.

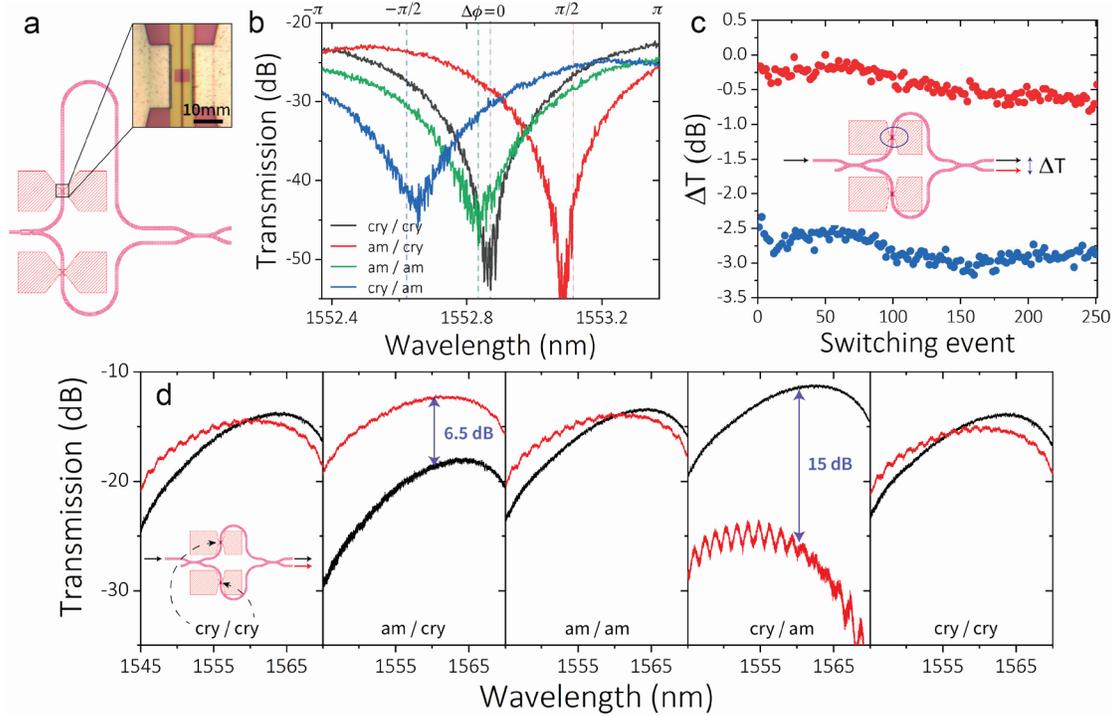

**Fig. 2. $Sb_2Se_3$-based phase shifters in balanced (2×2 switch) and unbalanced MZI devices. a** Layout of an unbalanced MZI with an optical microscope image of a 6 μm $Sb_2Se_3$ cell on a 10 μm-long microheater. **b** Experimental demonstration of four interferograms corresponding to each phase state permutation in the device shown in **a**. Crystallization was achieved using 6.2 V [3.2V] and 1 ms pulses. Amorphization was achieved with 21 V [10.8 V] and 400 ns pulses. Extinction ratio over 35 dB is achieved between opposite states with a nearly complete π phase shift modulation. **c** Reversible switching over 125 full cycles between two intermediate $Sb_2Se_3$ states of the top-arm phase shifter in a balanced MZI. The plotted changes in transmission correspond to the difference between each arm's intensity. **d** MZI 2×2 switch in the four different permutations: amorphous and crystalline states of each arm's phase shifter with 6 μm $Sb_2Se_3$ cell for π/2 shift. Reversible switching is demonstrated for both cells. The Gaussian-like transmission curves result from the use of grating couplers.

We now introduce $Sb_2Se_3$-based phase shifters in micro-ring resonators (Fig. 3a). Like MZI devices, we demonstrate continuous, multi-state modulation of an all-pass resonator by partially crystallizing a 3 μm-long $Sb_2Se_3$ cell (Fig. 3b). Figure 3c shows the reversible switching of the cell in 10 cycles between fully amorphous and fully crystalline states with remarkable reproducibility and excellent agreement with the theoretical value of 0.09 π/μm (estimated considering only the $Sb_2Se_3$ cell length). The losses upon crystallization were 0.018 dB/μm, which is quantified from fitting the transmission spectra of the two states (Supplementary Section S5). We then measured micro-ring resonators with an add-drop configuration and a 6 μm-long $Sb_2Se_3$ cell (Fig. 3d). Figure 3e and 3f demonstrate the phase-only modulation changing the resonance wavelength of the through-port dip or drop-port peak and the ~ π/2 phase modulation of the ultra-compact device.

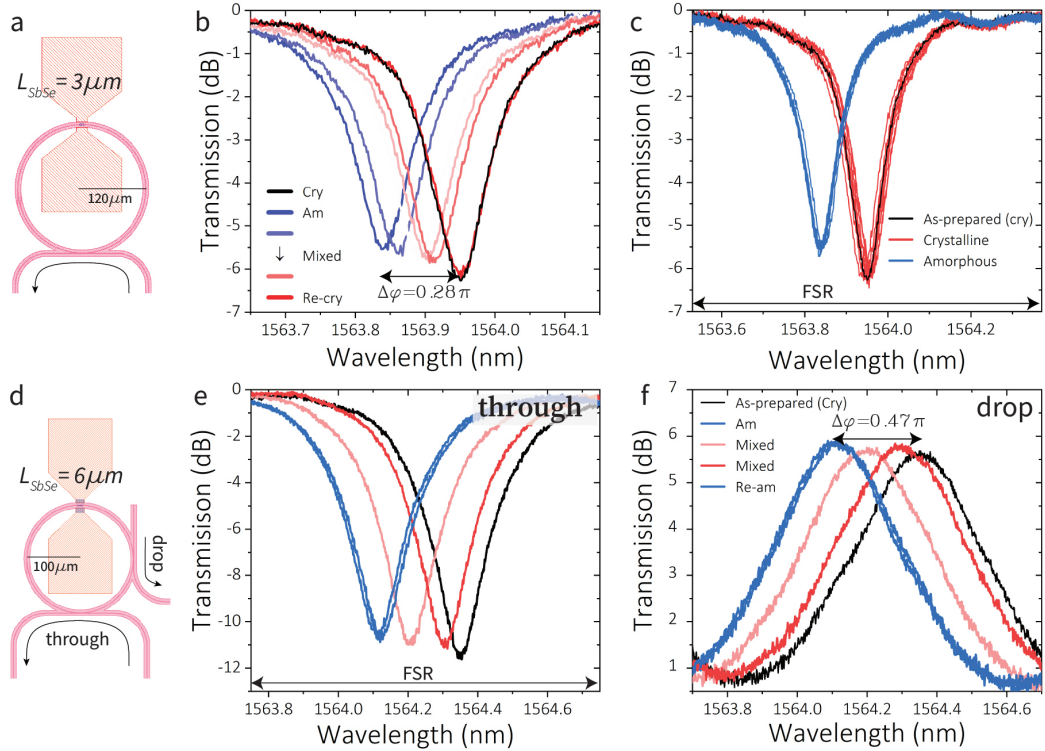

**Figure 3. Reversible and multi-level switching of Sb$_2$Se$_3$ phase shifters in micro-ring resonators. a** Layout of a 120 μm-radius, 200 nm-gap micro-ring resonator with a 3 μm-long Sb$_2$Se$_3$ cell. This device is used to experimentally demonstrate **b** one full reversible cycle featuring some of the partial crystallization states and a total of 0.28π shift, in excellent agreement with the expected 0.09 π/μm, and **c** ten full cycles with excellent reproducibility. **d** Same ring resonator as in **a** featuring a drop port identical to the main coupling port. **e** Demonstration of phase shift of resonance dip modulation in the through port with up to 10 dB in ER modulation. **f** Drop port peaks for the same exact measurements shown in **e** with a total phase shift of 0.47π. 3.2 V and 100 μs - 1ms pulses were used to partially crystallize and a single 21 V [10.8 V] and 400 ns pulse to fully amorphize.

In addition to spectral detuning of the resonant peak, we also demonstrate that the phase shifter can be used to modulate the extinction ratio of an over-coupled ring resonator by tuning the waveguide effective index at the coupling region, which modifies the coupling coefficient between the ring and the drop port[30] (Fig. 4a and Supplementary Section S7). We explored the two different configurations shown in Figs. 4b and 4c with Sb$_2$Se$_3$ cells covering either the ring waveguide only or both waveguides, respectively. Large extinction ratio modulation up to 20 dB was demonstrated in the latter configuration, where amorphization of the Sb$_2$Se$_3$ cell results in light lost to the drop port, therefore increasing the losses inside the ring, which then switches from over-coupled to near-critically-coupled response. As we will show later in this paper, simultaneous amplitude and phase trimming are potential applications of Sb$_2$Se$_3$-based phase shifters, allowing the correction of fabrication variations in the targeted resonance wavelength and in the intended coupling regime.

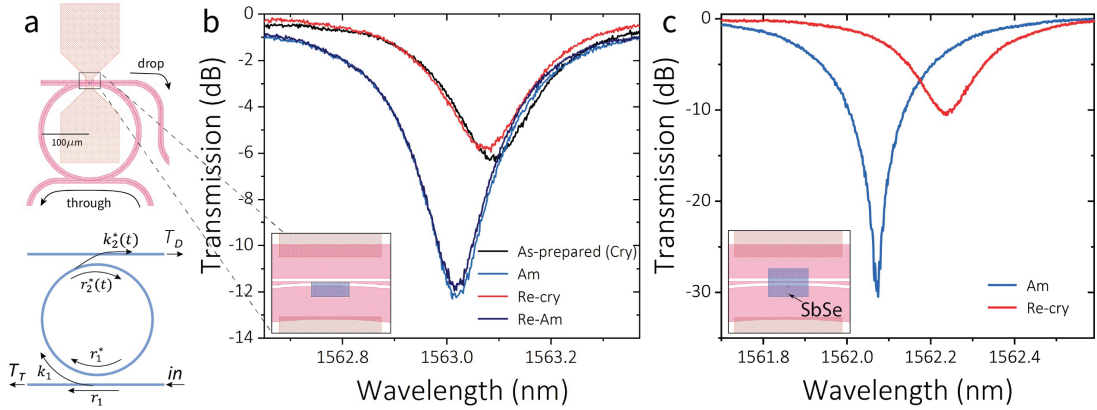

**Figure 4. Extinction ratio tuning via phase-shift triggered coupling coefficient modulation.** **a** Schematic of the ring resonators under test with a drop port at the phase shifter position and the corresponding modulation of the coupling coefficients *k* and *r*. **b** Experimental results for the device's through port for two switching cycles of a $Sb_2Se_3$ cell placed on the ring waveguide only, as shown in the inset. **c** Same as **b** for a $Sb_2Se_3$ cell covering both the ring waveguide and the drop port with nearly 20 dB modulation in ER.

Up to this point, we have shown $Sb_2Se_3$ phase shifters capable of achieving intermediate states via partial recrystallization only – modulating the number and the width of the crystallization pulses. As a result, when reconfiguring the OPS to a state with lower crystallinity, at least two pulses are needed, where one fully amorphizes and 'resets' the material while the other partially crystallizes the PCM. While this has so far been the standard procedure in PCM photonics for continuous electrical switching,[29,31,32] it not only complicates the switching scheme but also significantly compromises the switching speed and energy efficiency, given that crystallization pulses are inherently much longer and energy consuming than the amorphization pulse.[33] A switching approach that enables fast bidirectional tunability is therefore strongly preferred. Here we demonstrate a unique single-step partial amorphization scheme by engineering the temperature profile across a microheater. For this purpose, a microheater shown in Figs. 5a and 5b consisting of five *n*-doped 'bridges' each of 2 μm in width and connecting both $n^{++}$ regions was employed. $Sb_2Se_3$ is only deposited directly on top of each bridge. Changing the pulse width, the temperature profile along the waveguide and across the 'bridges' leads to melting temperatures initially only in the central 'bridge,' but then expanding to the outer ones as the pulse width (i.e., energy dissipation) increases. This response, in turn, allows for partial amorphization of the $Sb_2Se_3$ cells, correspondingly leading to different optical responses and phase shifting. With a total resistance of $1025 \pm 25$ Ω, the pulse width to start inducing amorphization was 800 ns with 21 V [10.8 V]. By varying the pulse width, *W*, from 800 to 1050 ns; we demonstrate partial amorphization of $Sb_2Se_3$ in an unbalanced MZI (Fig. 5c) and a micro-ring resonator (Fig. 5d). Figure 5c shows reversible switching between an intermediate state achieved with a 900 μs pulse and the fully crystalline state. Bowtie-shaped microheaters in Fig. 1 and Fig. 2 require 176 nJ for full amorphization followed by a partial crystallization in the 4 μJ - 38 μJ range if the target intermediate state was less crystalline (more amorphous). Using microheaters with 'bridges', these two pulses are replaced by a single partial amorphization pulse of 388 nJ, for a total of one to two orders of magnitude lower energy consumption per switching event. The concept of leveraging temperature gradient to achieve precisely controlled partial amorphization can be readily extended to other heater designs with more tuning levels.

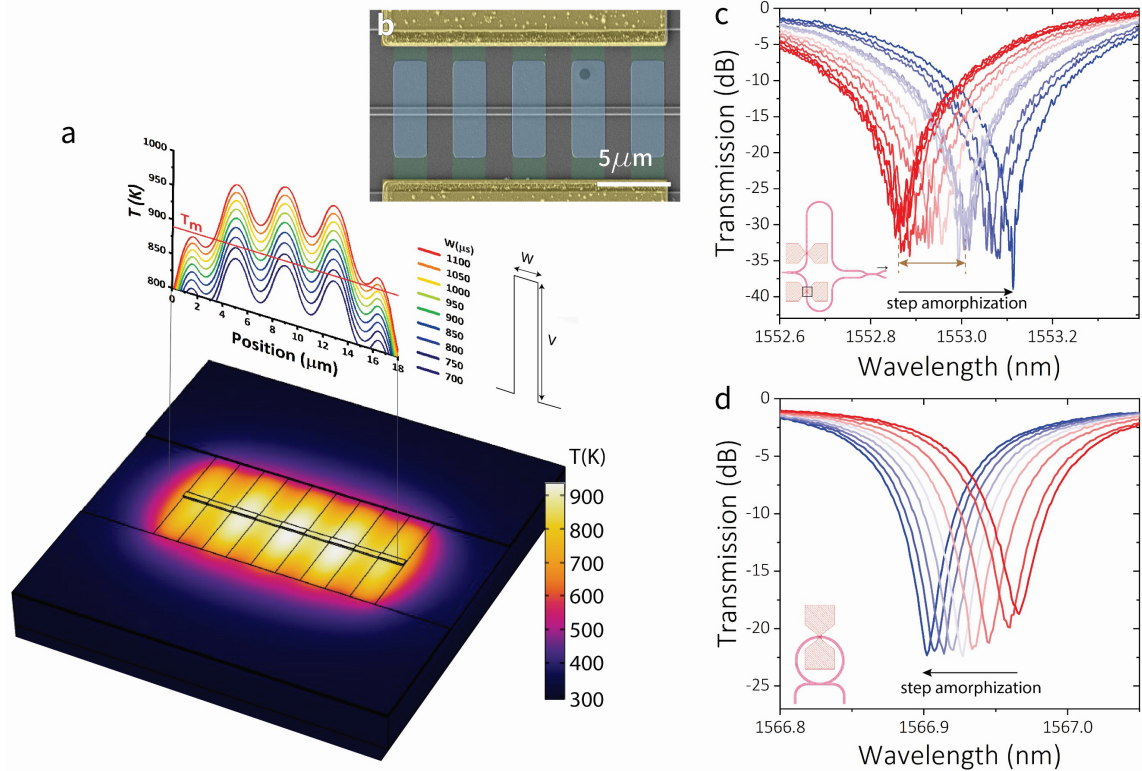

**Figure 5. Engineering microheater to achieve partial amorphization. a** FEM heat transfer simulation of a microheater consisting of five 2 μm-wide 'bridges'. The temperature profile along the waveguide is plotted for different pulse widths. **b** Colored SEM image of a fabricated device. **c** Partial amorphization of the bottom arm $Sb_2Se_3$ in an unbalanced MZI. 10.8 V pulses with varying widths between 800 ns and 1050 ns were employed to amorphize and 3.2 V for 1 ms to fully crystallize. The transmission is measured in the top output waveguide where a maximum of π/2 phase shift is observed, corresponding to the full amorphization of the three central 'bridges'. Reversible switching is demonstrated over three cycles between the full crystalline state and the marked intermediate state (corresponding to a 900 ns pulse). **d** The same phase shifter with partial amorphization in a micro-ring resonator with continuous tuning of the resonance wavelength.

Leveraging the bidirectional tuning capability, we lastly demonstrate an application that exploits the nonvolatile nature of chalcogenide PCMs: optical device trimming. We use a device comprising two micro-ring resonators connected to a single bus waveguide (Fig. 6) to show the potential of PCMs in countering the optical response variations resulting from fabrication non-uniformity. By tuning the phase of one of the micro-ring resonators and keeping the second as a reference, we demonstrate that a nonvolatile phase shift can be used to either isolate each ring's response (Fig. 6a) or to make them coincide by precisely tuning the resonance wavelength until matching (Fig. 6b). Furthermore, Figure 6c shows how both amorphization and crystallization processes can be combined to shift the peak position in both directions and correct any undesired over-shift if the goal is for the peaks to match. Once the trimming is completed, and thanks to the well-known nonvolatility of PCMs,[17,34] the final optical response is held stable over time, as we demonstrate up to $4 \times 10^6$ s (i.e. 47 days) later, with a small shift of 0.01 nm, which is difficult to differentiate from that associated with temperature fluctuations (see Supplementary Information). More importantly, our low-loss phase shifter still offers the option of recalibrating or changing the configuration later without any power consumption in between.

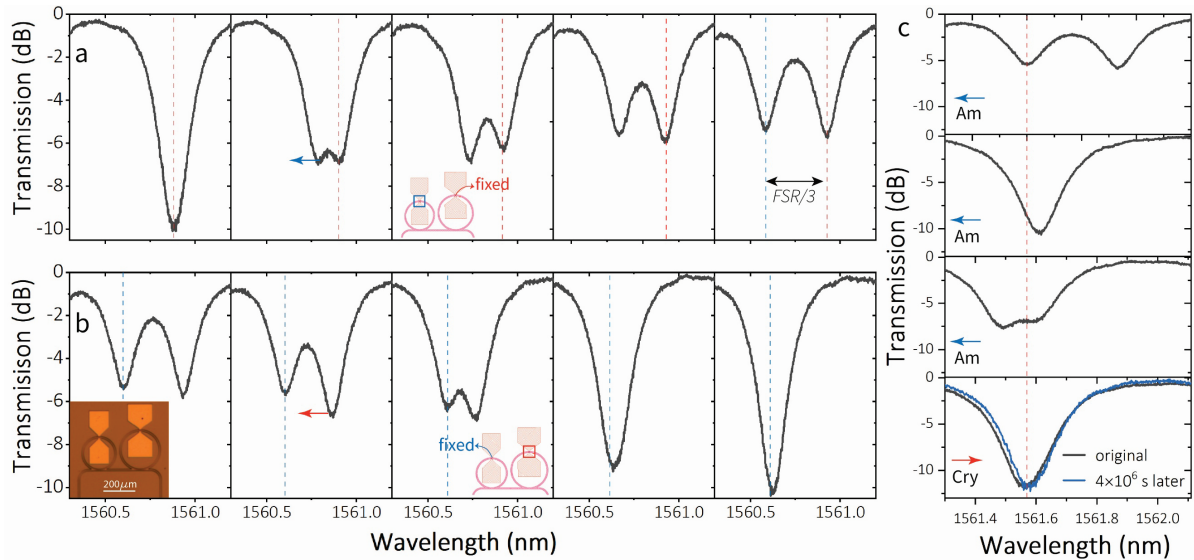

**Figure 6. Optical trimming of micro-ring resonators. a** Detuning of two initially superimposed resonance dips corresponding to two micro-ring resonators with 100 μm (blue line) and 120 μm (red line) in radius connected to a common bus waveguide. The largest ring is left fixed while the smaller ring is phase-shifted via step amorphization to fully isolate the response of each ring. **b** Trimming of the 120 μm micro-ring to match the resonance dip of the smaller ring (fixed), thus, correcting the detuning in **a** but with an overall wavelength shift of 0.6 nm. The inset shows an optical microscope image of the actual device. **c** Demonstration of combined amorphization and crystallization to properly trim one micro-ring resonator and compensate 'over-shifting'. Nonvolatile response is demonstrated in the last state of the device (bottom panel) which was measured $4 \times 10^6$ s (i.e. 47 days) later.

**Discussion**

We have demonstrated electrothermal control of the low-loss phase change material $Sb_2Se_3$ as a nonvolatile phase shifter in various reconfigurable photonic devices. Our choice of phase change materials is based on the large bandgap and the fact that $Sb_2Se_3$ is a single-phase compound throughout the entire solid-liquid temperature range, which, in principle, means high endurance. Considering that electrical current does not flow through the material in our approach, electromigration is also prevented, which is a mechanism for atomic movement and failure. In this work, we tested the endurance of over 250 switching events, a number ultimately limited by our experimental setup; however, 4,000 switching cycles have been demonstrated using laser pulses.[27]

From simulations, we expected a phase shift of 0.09 π/μm, which was in excellent agreement with the experimental demonstrations shown for smaller Sb$_2$Se$_3$ cells (Fig. 1). For longer cells, the total phase modulation was slightly smaller, down to 0.08 π/μm. We attribute this discrepancy to possible thickness variations of the Sb$_2$Se$_3$ film and non-uniformity of the heating profile. Nonetheless, we demonstrated that an ultra-compact device comprising only a 10 μm-long microheater and a 6 μm-long Sb$_2$Se$_3$ is enough to introduce π/2 phase shift. In turn, two π/2 nonvolatile phase shifters in a balance MZI switch, one per arm, allowed us to demonstrate another fundamental building block of programmable photonic circuits, a 2 × 2 optical switch (Fig. 2). We have also demonstrated the phase tunability of all-pass and add-drop micro-ring resonators to control the signals in the through and drop ports (Fig. 3), and amplitude modulation by engineering the coupling efficiencies to a drop port (Fig. 4). The insertion loss of the phase shifter (0.03 dB/μm) can be lowered by further engineering the shape of the PCM cell to suppress scattering.[35] Additionally, the microheater doping profiles and geometry can also be optimized. A PIN configuration, which has demonstrated losses of 0.02 dB/μm[32] can be a good alternative. Other alternatives for transparent microheaters, already demonstrated to switch PCMs, might be considered to avoid doping losses overall and extend the transparency window to shorter wavelengths, such as graphene,[36] fluorine-doped tin oxide,[37] and indium-tin-oxide;[35,38] however, their full integration and reversible cyclability on integrated waveguides is yet to be demonstrated. We draw a direct comparison between our PCM approach and other phase-shifting approaches in Table 1 considering 10%-90% rise time, the total change in the effective index, $\Delta n_{eff}$, the total length and voltage to achieve a π phase shift, $L_\pi$ and $V_\pi$, and the insertion loss (IL).

The nonvolatility of our phase shifter guarantees zero-static power consumption – a new paradigm for PICs. Even if reconfigured every millisecond, our approach's energy efficiency is still superior to TO phase shifters. However, if we consider only applications in which programming is done sporadically, then the energy savings are orders of magnitude different. For example, let us consider a 32 × 32 optical switching fabric architecture,[22] consisting of 512 2 × 2 switches for a total of 1024 OPS (using the design in Fig. 2d), whose output is reconfigured every 1 s. Using our Sb$_2$Se$_3$-based phase shifters, the average switching energy – calculated using the full crystallization pulse, which is the most energy-consuming process – would be 39.3 mJ. This is two orders of magnitude lower than a constant supply of 1.9 J to reconfigure the lowest reported 32 × 32 network with the same MZI switches but using TO phase shifters.[39] The energy-saving benefit becomes even more significant with longer durations between switching events. Such nonvolatility is also interesting for in-memory computing,[34,40–42] and configure-once applications, such as optical trimming and transient coupler for wafer-scale photonic testing.[43] Moreover, in addition to the nonvolatile modulation by phase transitions of Sb$_2$Se$_3$, the doped-silicon microheater can provide conventional thermo-optical modulation (see supplementary information), which further enhances the versatility of our approach by enabling dual nonvolatile and volatile response in a single device. We note that while the contact pads to electrically connect our devices are >1 mm$^2$, they do not represent a limitation in terms of scalability. In

3D stacking architectures, vertical vias can be used to contact the doped-silicon while keeping on-silicon footprints of 20×20μm².

**Table 1.** Comparison of phase shifter platforms currently available in 220 nm SOI at 1550 nm. $E_\pi$ is the energy required to switch and hold the state. ΔT is the difference in temperature, and t[s] is the time of operation.

| | $\Delta n_{eff}$ | IL(dB) | $L_\pi$ (μm) | $E_\pi$ | $V_\pi$ (V) | Non-volatility | Rise time[d] |
|---|---|---|---|---|---|---|---|
| Thermo-optical (doped-Si)[12] | ~1.8 × 10⁻⁴ ×ΔT[K] | 0.23 | 61.6 | 24.8 × t[s] mJ | 4.36 | No | 2.69 μs |
| Thermo-optical (metal heater)[12,44] | ~1.8 × 10⁻⁴ ×ΔT[K] | < 0.4 | >200 | 20-25 × t[s] mJ | 4 | No | 3.7 μs |
| Thermo-optical (ITO)[45] | ~1.8 × 10⁻⁴ ×ΔT[K] | <0.01 | 50 | 10 × t[s] mJ | 9.1 | No | 5.2 μs |
| Thermo-optical with slow light[46] | ~0.1 | 2 | ~10 | 2 × t[s] mJ | 4.5 | No | 100 ns |
| Electro-optical (depletion)[47] | 1.5 × 10⁻⁴ | 3.6 | 1500 | ~2 × t[s] pJ [a] | 10.7 | No | 15.5 ps |
| Electro-optical (injection)[48] | ~10⁻³ | 2 | 400 | 1.7 × t[s] mJ | 2 | No | 24 ns |
| Optoelectro-mechanical[13,15] | ~0.01 | 0.47 | 84 | 179 × t[s] nJ | 2 | No | 123 ns |
| Liquid crystal[9] | 0.016 | 0.25 | 49 | 1 × t[s] nJ | 5 | No | 1 ms |
| Plasmonic nonlinear polymer[10c] | 0.055 | 12 | 29 | 4 × t[s] mJ | 45 | No | 15 ps |
| BTO Pockels effect[11] | 7 × 10⁻⁴ | 1 | 1000 | 8.2 × t[s] mJ | 3.3 | No | 11.7 ps |
| BTO Ferroelectric domain switching[16] | ~1 × 10⁻⁴ / pulse | N/A[e] | 150 | N/A[e] | -10 | Yes | 300 ns |
| LiNbO₃ [49] | 1.5 × 10⁻⁴ | 2.5 | 5000 | 53 × t[s] mJ | 5.1 | No | 3.5 ps |
| Transition metal dichalcogenides[50] | 1.5× 10⁻³ | 0.55 | 1000 | 0.64 × t[s] nJ[a] | 8 | No | 1 ns |
| 30 nm Sb₂Se₃ cladding cry/am (this work) | 0.07 | 0.36[b] | 11 | 38.4 μJ / 176 nJ | 6.2/ 21.0 | Yes | 0.1-1 ms / 800 ns |

[a] No current flows in this device, making the power dissipation extremely low. However, a constant supply of large voltage means a significant power consumption of the electronics. [b] Based on a 12 μm-long Sb₂Se₃ OPS – it can be further decreased if using lower doping concentrations. [c] These values are calculated from the supplementary material of the referenced paper. [d] 10%-90% rise time, where we used the single-pole approximation: $\tau = 0.35/f_{3dB}$ for results with no rise time reported. [e] No information available, although BTO is transparent at telecom wavelengths and no current flows through the device. However, the modulation depends on large number of pulses.

In conclusion, we have experimentally demonstrated a new class of electrically-driven phase shifter that exploits the unique optical properties of phase-change materials to achieve zero-static power consumption in ultra-compact devices. We demonstrated several active photonic devices and applications as well as the potential of these devices for low-energy programmable photonic integrated circuits. Our results are promising for applications in optical trimming to correct fabrication variations and in photonic architectures with sporadic reprogramming, such as optical switching fabrics for telecommunications, phased arrays for particle trapping, photonic computing (inference with static matrices, for instance), and quantum networks, among others.

**Methods**

*Device Fabrication*

Ten 220 nm SOI wafers were fabricated with varying *n*-doping concentrations using the 90-nm CMOS line in Lincoln Laboratory's 200 mm wafer foundry (Supplementary Section S1). The doping of silicon was carried out using ion implantation of phosphorous with varying dose and implantation energy. The $n^{++}$ region was formed with a dose of $10^{16}$ cm$^2$ and an ion energy of 80 keV. The *n* region of ~$4\times10^{18}$ cm$^{-3}$ was formed with a dose of $10^{14}$ cm$^{-2}$ and an ion energy of 80 keV. The contact was fabricated with a Ti/TiN barrier following a metallization with aluminum and passivation with $SiO_2$. 500 nm-wide SOI waveguides and $Sb_2Se_3$ cells were fabricated following two electron beam lithography fabrication steps on an Elionix ELS-F125 system (Supplementary Section S1). The waveguides were patterned using ZEP 520A positive photoresist followed by chlorine etching of half the silicon thickness. A lift-off process with polymethyl methacrylate (PMMA) resist was used to open the windows for the subsequent thermal evaporation of 30 nm $Sb_2Se_3$. The film deposition was performed using thermal evaporation from $Sb_2Se_3$ bulk materials following previously established protocols.[51,52] Bulk starting $Sb_2Se_3$ was synthesized using a standard melt-quench technique from high-purity (99.999%) raw elements.[53] The samples were annealed in an argon environment at 200 °C for 10 – 15 min before depositing 15 nm of $Al_2O_3$ by atomic layer deposition to obtain a conformal protective layer.

*Experimental setup*

A home-built wafer-scale automated photonic testing system was used to perform the optoelectronic measurements. A custom 18-channel SMF-28 Ultra Fiber Array with a 250 µm pitch (PLC-Connections) was used to couple light in and out of the chip using silicon-etched grating couplers. A C+L band optical vector analyzer (LUNA OVA5000) was used to scan the spectrum and conduct time-domain measurements. The high voltage electrical pulses were generated with an analog 1 GHz, 40 V pulse generator (E-H Research 136A) with a minimum of 0.5 ns raising and trailing edges, and the crystallization pulses with an Agilent 33250A 100MHz pulse generator. An ACP40-AW-GSG-250 air-coplanar probe (FormFactor) was used to contact the electrical pads on the chip.

*Microheater modeling*

The Joule heating process and heat dissipation model were performed using a three-dimensional finite-element method simulation in COMSOL Multiphysics. We used the Semiconductors module to simulate the carrier mobility in doped silicon with the same values and shapes as our fabricated devices. We coupled this module with the Heat Transfer in Solids, where surface-to-surface radiation and thermal boundary resistance were considered. The temperature in Heat Transfer was coupled back to the semiconductor module. The mode calculations shown in Fig. 1d were carried out in Lumerical Mode, considering the $Sb_2Se_3$ refractive index shown in Fig. 1b and the built-in refractive indices for Si, $Al_2O_3$, and $SiO_2$.

## Acknowledgments

This material is based upon work supported by the DARPA Young Faculty Award Program under Grant Number D18AP00070, the Assistant Secretary of Defense for Research and Engineering under Air Force Contract No. FA8702-15-D-0001, and Draper Laboratory. The authors acknowledge fabrication facility support by the MIT Microsystems Technology Laboratories, the Materials Research Laboratory, and the Lincoln Laboratory Microelectronic Laboratory.

## Data Availability

The data that support the findings of this study are available from the corresponding authors upon reasonable request.

## References

1. Bogaerts, W. *et al.* Programmable photonic circuits. *Nature* **586**, 207–216 (2020).
2. Carolan, J. *et al.* Universal linear optics. *Science* **349**, 711–716 (2015).
3. Cheng, Q., Bahadori, M., Glick, M., Rumley, S. & Bergman, K. Recent advances in optical technologies for data centers: a review. *Optica* **5**, 1354 (2018).
4. Sun, J., Timurdogan, E., Yaacobi, A., Hosseini, E. S. & Watts, M. R. Large-scale nanophotonic phased array. *Nature* **493**, 195 (2013).
5. Kita, D. M. *et al.* High-performance and scalable on-chip digital Fourier transform spectroscopy. *Nature Communications* **9**, 1–7 (2018).
6. Shen, Y. *et al.* Deep learning with coherent nanophotonic circuits. *Nature Photonics* **11**, 441–446 (2017).
7. Arrazola, J. M. *et al.* Quantum circuits with many photons on a programmable nanophotonic chip. *Nature* **591**, 54–60 (2021).
8. Witzens, J. High-Speed Silicon Photonics Modulators. *Proceedings of the IEEE* **106**, 2158–2182 (2018).
9. Pfeifle, J., Alloatti, L., Freude, W., Leuthold, J. & Koos, C. Silicon-organic hybrid phase shifter based on a slot waveguide with a liquid-crystal cladding. *Optics Express* **20**, 15359 (2012).
10. Melikyan, A. *et al.* High-speed plasmonic phase modulators. *Nature Photonics* **8**, 229–233 (2014).
11. Abel, S. *et al.* Large Pockels effect in micro- and nanostructured barium titanate integrated on silicon. *Nature Materials* **18**, 42–47 (2019).
12. Harris, N. C. *et al.* Efficient, compact and low loss thermo-optic phase shifter in silicon. *Optics Express* **22**, 10487 (2014).
13. Grottke, T., Hartmann, W., Schuck, C. & Pernice, W. H. P. Optoelectromechanical phase shifter with low insertion loss and a $13\pi$ tuning range. *Optics Express* **29**, 5525 (2021).
14. Grajower, M., Mazurski, N., Shappir, J. & Levy, U. Non-Volatile Silicon Photonics Using Nanoscale Flash Memory Technology. *Laser and Photonics Reviews* **12**, 1–8 (2018).
15. Errando-Herranz, C. *et al.* MEMS for Photonic Integrated Circuits. *IEEE Journal of Selected Topics in Quantum Electronics* **26**, (2020).
16. Geler-Kremer, J. *et al.* A Non-Volatile Optical Memory in Silicon Photonics. *2021 Optical Fiber Communications Conference and Exhibition, OFC 2021 - Proceedings* 12–14 (2021).
17. Wuttig, M., Bhaskaran, H. & Taubner, T. Phase-change materials for non-volatile photonic applications. *Nature Photonics* **11**, 465–476 (2017).
18. Abdollahramezani, S. *et al.* Tunable nanophotonics enabled by chalcogenide phase-change materials. *Nanophotonics* **9**, 1189–1241 (2020).

19. Wang, J., Wang, L. & Liu, J. Overview of Phase-Change Materials Based Photonic Devices. *IEEE Access* **8**, 121211–121245 (2020).
20. Rios, C. *et al.* Integrated all-photonic non-volatile multi-level memory. *Nature Photonics* **9**, 725–732 (2015).
21. Zhang, H. *et al.* Ultracompact Si-GST hybrid waveguides for nonvolatile light wave manipulation. *IEEE Photonics Journal* **10**, 1–10 (2018).
22. Zhang, Q. *et al.* Broadband nonvolatile photonic switching based on optical phase change materials: beyond the classical figure-of-merit. *Optics Letters* **43**, 94 (2018).
23. Michel, A. K. U. *et al.* Using low-loss phase-change materials for mid-infrared antenna resonance tuning. *Nano Letters* **13**, 3470–3475 (2013).
24. Dong, W. *et al.* Wide Bandgap Phase Change Material Tuned Visible Photonics. *Advanced Functional Materials* **29**, 1806181 (2019).
25. Liu, H. *et al.* Rewritable color nanoprints in antimony trisulfide films. *Science Advances* **6**, 7171–7187 (2020).
26. Delaney, M., Zeimpekis, I., Lawson, D., Hewak, D. W. & Muskens, O. L. A New Family of Ultralow Loss Reversible Phase-Change Materials for Photonic Integrated Circuits: $Sb_2S_3$ and $Sb_2Se_3$. *Advanced Functional Materials* **30**, 2002447 (2020).
27. Delaney, M. *et al.* Nonvolatile programmable silicon photonics using an ultralow-loss $Sb_2Se_3$ phase change material. *Science Advances* **7**, 1–8 (2021).
28. Faneca, J. *et al.* Towards low loss non-volatile phase change materials in mid index waveguides. *Neuromorphic Computing and Engineering* **1**, 014004 (2021).
29. Zhang, H. *et al.* Miniature Multilevel Optical Memristive Switch Using Phase Change Material. *ACS Photonics* **6**, 2205–2212 (2019).
30. Li, Y. *et al.* Coupled-ring-resonator-based silicon modulator for enhanced performance. *Optics Express* **16**, 13342 (2008).
31. Zhang, Y. *et al.* Electrically reconfigurable non-volatile metasurface using low-loss optical phase-change material. *Nature Nanotechnology* **16**, 661–666 (2021).
32. Zheng, J. *et al.* Nonvolatile Electrically Reconfigurable Integrated Photonic Switch Enabled by a Silicon PIN Diode Heater. *Advanced Materials* **32**, 2001218 (2020).
33. Zhang, Y. *et al.* Myths and truths about optical phase change materials: A perspective. *Applied Physics Letters* **118**, (2021).
34. Ríos, C. *et al.* In-memory computing on a photonic platform. *Science Advances* **5**, eaau5759 (2019).
35. Taghinejad, H. *et al.* ITO-based microheaters for reversible multi-stage switching of phase-change materials: towards miniaturized beyond-binary reconfigurable integrated photonics. *Optics Express* **29**, 20449 (2021).
36. Ríos, C. *et al.* Multi-Level Electro-Thermal Switching of Optical Phase-Change Materials Using Graphene. *Advanced Photonics Research* **2**, 2000034 (2020).
37. Youngblood, N. *et al.* Reconfigurable Low-Emissivity Optical Coating Using Ultrathin Phase Change Materials. *ACS Photonics* **9**, 90–100 (2022).
38. Kato, K., Kuwahara, M., Kawashima, H., Tsuruoka, T. & Tsuda, H. Current-driven phase-change optical gate switch using indium-tin-oxide heater. *Applied Physics Express* **10**, (2017).
39. Suzuki, K. *et al.* Low-Insertion-Loss and Power-Efficient 32 × 32 Silicon Photonics Switch with Extremely High-Δ Silica PLC Connector. *Journal of Lightwave Technology* **37**, 116–122 (2019).
40. Feldmann, J. *et al.* Parallel convolution processing using an integrated photonic tensor coer. *Nature* vol. 589 52–58 (2021).
41. Wu, C. *et al.* Programmable phase-change metasurfaces on waveguides for multimode photonic convolutional neural network. *Nature Communications* **12**, 1–8 (2021).
42. Shastri, B. J. *et al.* Photonics for artificial intelligence and neuromorphic computing. *Nature Photonics* **15**, 102–114 (2021).
43. Zhang, Y. *et al.* Transient Tap Couplers for Wafer-Level Photonic Testing Based on Optical Phase Change Materials. *ACS Photonics* **8**, 1903–1908 (2021).
44. Jacques, M. *et al.* Optimization of thermo-optic phase-shifter design and mitigation of thermal crosstalk on the SOI platform. *Optics Express* **27**, 10456 (2019).
45. Parra, J., Hurtado, J., Griol, A. & Sanchis, P. Ultra-low loss hybrid ITO/Si thermo-optic phase shifter with optimized power consumption. *Optics Express* **28**, (2020).
46. Vlasov, Y. A., O'Boyle, M., Hamann, H. F. & McNab, S. J. Active control of slow light on a chip with photonic crystal waveguides. *Nature* **438**, 65–69 (2005).

47. Zhang, F. *et al.* Toward single lane 200G optical interconnects with silicon photonic modulator. *Journal of Lightwave Technology* **38**, 67–74 (2019).
48. Kang, G. *et al.* Silicon-Based Optical Phased Array Using Electro-Optic Phase Shifters. *IEEE Photonics Technology Letters* **31**, 1685–1688 (2019).
49. He, M. *et al.* High-performance hybrid silicon and lithium niobate Mach–Zehnder modulators for 100 Gbit s -1 and beyond. *Nature Photonics* **13**, 359–364 (2019).
50. Datta, I. *et al.* Low-loss composite photonic platform based on 2D semiconductor monolayers. *Nature Photonics* **14**, (2020).
51. Zhang, Y. *et al.* Broadband transparent optical phase change materials for high-performance nonvolatile photonics. *Nature Communications* **10**, 2–3 (2019).
52. Hu, J. *et al.* Fabrication and testing of planar chalcogenide waveguide integrated microfluidic sensor. *Optics Express* **15**, 2307 (2007).
53. Petit, L. *et al.* Compositional dependence of the nonlinear refractive index of new germanium-based chalcogenide glasses. *Journal of Solid State Chemistry* **182**, 2756–2761 (2009).